# Prandtl number of lattice Bhatnagar-Gross-Krook fluid

Y. Chen, H. Ohashi, and M. Akiyama

*Department of Quantum Engineering and Systems Science, University of Tokyo*

**Abstract**

The lattice Bhatnagar-Gross-Krook modeled fluid has an unchangeable unit Prandtl number. A simple method is introduced in this letter to formulate a flexible Prandtl number for the modeled fluid. The effectiveness was demonstrated by numerical simulations of the Couette flow.

Lattice Boltzmann (LB) method[1], which is a semi-continuous version of lattice gas automata (LGA)[2], has been accepted recently as an alternative numerical tool for the simulation of transport phenomena in the fluid system. The latest modification[3-5] of the LB method does remove many disadvantages of the anecstral LGA model, such as the lack of Galilean invariance, the velocity dependent pressure, the high noise-signal ratio, the high viscosity and the inflexible sound speed, etc. On the other hand, the inheritated particle nature keeps the high efficiency for the LB method when applied in the simulation of flows under complicated geometries and reserves it as one of the most appealing candidates for the computing models on the massively parallel frame in the next decade.

Selecting a regular spatial lattice and rendering each link vector of this lattice into the vector for the particle flight velocity, a phase space for the particle motion can be defined. The single particle distribution may be denoted, in the discrete spatial and velocity spaces, as $N_{pi}(\vec{x},t)$, where $x_\alpha$ and $t$ are spatial and temporal coordinates. Indices, $p$ and $i$, are for sub-lattices and lattice links respectively. The lattice Boltzmann equation (LBE) to be solved in the process of the LB simulation can be written as,

$$N_{pi}(\vec{x}+\vec{c}_{pi}, t+1) - N_{pi}(\vec{x}, t) = \triangle_{pi}. \tag{1}$$

Here, $\vec{c}_{pi}$ is the link vector, which represents the flight velocity of particle in the meantime. The lhs of (1) stands for the propagation of particle, while $\triangle_{pi}$ is defined as the collision term, which is generically linearized as a product of a collision matrix and the perturbative part of particle distribution, i.e. $(N_{pi} - N_{pi}^{eq})$. The structure and elements of the collision matrix shall certainly ensure the conservation of densities of mass ($\rho$), momentum ($\rho\vec{u}$) and thermal energy ($e$). More often, the famous Bhatnagar-Gross-Krook (BGK) or the single time relaxation approximation (STRA) is introduced in (1), which transforms the collision term as simple as the following expression,

$$\triangle_{pi} = -\frac{1}{\tau}\left(N_{pi} - N_{pi}^{eq}\right). \tag{2}$$

Here $\tau$ is the relaxational time period, through which the particle distribution approaches its equilibrium value. When the equilibrium particle distribution is expanded in the low speed limit as

$$N_{pi} = A_p(\rho,e) + M_p(\rho,e)(u_\alpha c_{pi\alpha}) \tag{3}$$
$$+ G_p(\rho,e)u^2 + J_p(\rho,e)(u_\alpha c_{pi\alpha})^2 + \mathcal{O}(u^3),$$

and a set of constraints for the expansion parameters $A_p, M_p \ldots J_p$ is properly chosen, a perturbative multi-scale analysis would lead to macro-dynamic equations in the compressible Navier-Stokes form, provided that the lattice is symmetric enough to ensure the isotropy for the flux tensors appearing in these equations. This serves as a theoretical proof for the use of the LB or the lattice BGK fluid dynamic model and a justification for naming such a semi-continuous, interacting system the lattice BGK fluid, as in the title of this paper. The state equation of the current fluid is identical to that of perfect gas, because there is no inter-particle force considered. Hence the temperature is proportional to the thermal energy, specifically in two dimensions, one has $T = e$. Transport properties can be described once the shear viscosity and the heat conductivity are identified. Indeed, in the $D$ dimensional space they are written as follows,

$$\nu = \frac{2}{D}\rho e \left(\tau - \frac{1}{2}\right), \tag{4}$$

$$\kappa = \frac{D+2}{D}\rho e \left(\tau - \frac{1}{2}\right). \tag{5}$$

As in usual hydrodynamic text, $\nu$ is used for the shear viscosity and $\kappa$ the heat conductivity. The ratio of these two quantities will be independent of any thermodynamic variable except the space dimension, from both (4) and (5). Therefore, the Prandtl number of the lattice BGK fluid defined as $\frac{c_p\nu}{\kappa}$, where $c_p$ denotes the specific heat, would have a sole and unit numerical value.

The reason for the unit Prandtl number of the lattice BGK fluid may be understood through a more intuitive way: the form of the collision term in this model (2) indicates an incorporation of the diagonal matrix $-\frac{1}{\tau}\delta_{ij}$, where $\delta_{ij}$ is the Kronecker tensor. Eigenvalues for the shear and energy transport modes are equal in this case and therefore result in the unit value for the Prantl number.

The frozen unit-value Prandtl number for the lattice BGK fluid is not favorite, in sense of the flexibility of a fluid model, as in many cases one is interested in observing the change in the transport of momentum and heat, along with the variation of the Prandtl number. For example, a bifurcation from the steady roll solution to the time dependent thermal plumes is found in the high Rayleigh number flow in the natural convection, when the Prandtl number of the fluid switches from a lower to a higher numerical value.

To solve this difficulty in the lattice BGK model, an additional parameter $\sigma$ can be included in the aforementioned constraint for the velocity expansion parameter, specifically for $A_p$ in the following form,

$$\sum_{pi} c_{pi}^4 A_p = \frac{4(D+2)}{D^2} \sigma \rho e^2 \,. \tag{6}$$

With a standard derivation, (6) will make the ratio of heat and momentum transfer tunable, which can result in a $\sigma$ parameterized Prandtl number as follows,

$$\Pr = \frac{1}{2\sigma - 1} \,. \tag{7}$$

To verify this formula, numerical measurements of the Prandtl number have to be carried out. In this paper, the measurements were realized by simulating Couette flows between two parallel flat walls of which one is set at rest, the other moving with a constant velocity $U$ in its own plane. It is also postulated that the temperature is constant along the wall, but different for the two walls,

$$y = 0 : T = T_0; \quad y = h : T = T_1 \,. \tag{8}$$

Here $h$ is the distance between the two walls and in this case temperatures are set to be $T_0 \leq T_1$. When the flow reaches the steady stage, the temperature profile is measured and compared with the theoretical solution:

$$\frac{T - T_0}{T_1 - T_0} = \frac{y}{h} + \frac{\nu U^2}{2\kappa(T_1 - T_0)} \frac{y}{h}\left(1 - \frac{y}{h}\right) \tag{9}$$

$$= \eta + \frac{1}{2}(\Pr \times \mathrm{E})\eta(1-\eta) \,,$$

where $\eta = \frac{y}{h}$ is the normalized width for the channel and E is the dimensionaless Eckert number defined as $\frac{U^2}{c_p \Delta T}$. For $T_1 = T_0$ case, the solution is obtained as

$$T - T_0 = \frac{\nu U^2}{2\kappa} \eta(1-\eta) \,, \tag{10}$$

which has a parabolic profile across the channel.

The results are shown in Fig. 1. The quality of the modification in (6) is estimated by inspecting the devation between the meaured results and analytically predicted cureve. It is obvious from the figure that the current method is only reliable at the neighbourhood $\sigma = 1$. In other words, the Prandtl number can only be varied little from the coherent unit value. The reason for this failure can be disclosed by a further study of the heat flux tensor, which, because of the incorporation of $\sigma$, changed into the following form,

$$q_\alpha = \frac{4(D+2)}{D^2} \tag{11}$$

$$[(2\sigma - 1)\rho e \partial_\alpha e + (1-\sigma)e^2 \partial_\alpha \rho]\left(\tau - \frac{1}{2}\right) \,.$$

The existence of $\partial_\alpha \rho$ is unphysical and causes the disagreement stated above. In order to do away with the effect of the density gradient term, another constraint involving $\sigma$ may be considered as follows,

$$\sum_{pi} c_{pi}^4 A_p = \frac{4(D+2)}{D^2} \rho[(1-\sigma)\bar{e}^2 + \sigma e] \,. \tag{12}$$

where $\bar{e}$ is the averaged termperature. The unphysical part of the heat flux tensor then becomes

$$(1-\sigma)(\bar{e}^2 - e^2)\left(\tau - \frac{1}{2}\right)\partial_\alpha \rho \,, \tag{13}$$

so that it has the minimum effect when the distribution of thermal energy in the flow is uniform enough, namely in the small temperature difference limit, provided that $\sigma$ is not too far from the unit value.

The effectiveness of such a formulation can be demonstrated by the same numerical measurement of the Prandtl number described before. Again results are plotted in Fig. 1 with different symbols for different temperature differences. The agreement with (7) is extremely well, when $\sigma$ is selected from 0.65 $\sim$ 1.35. Correspondingly the Prandtl number may vary from 3.3 to 0.59. The breakdown of this method had been also observed either when $\sigma$ exceeds the range, or when the temperature difference is too large, e.g. $\Delta e \geq 0.1$. These may be considered as restrictions when the new lattice BGK model with variable Prandtl number is to be applied in any thermohydraulic problems.

First application of this model is chosen as the Couette flow used above as the numerical experiment to measure the Prandtl number of the modeled fluid. Temperature profiles in steady state are shown in Fig. 2. The Eckert numbers for these simulations are the same, i.e. E = 2.0, and the temperature difference on the two walls is set to 0.01, keeping the method under the valid condition. Subsequently different values for $\sigma$ are provided to vary the Prandtl number of the fluid as 0.5, 1.0, 2.0. As the product Pr × E changes, the heat flux from the upper wall to the fluid changes sign. From the temperature profiles in Fig. 2, it is clear that cooling of the upper wall occurs in case of Pr × E < 2, heating of the upper wall in case of Pr × E > 2, and no heat flows between the upper wall and the lattice BGK fluid as Pr × E = 2. Furthermore, numerical results and the Naver-Stokes solution are found to remarkably agree with each other in the same figure.

Change of the Prandtl number of a fluid system in quantitative sense by microscopic approaches is nontrivial, as even for the full and continuous Boltzmann equation, it is difficult to find the corresponding collisional cross sections[6]. In the case of discrete Boltzmann equation, collision matrices which have different eigenvalues controlling transport modes for heat and momentum can be used[7] to solve the difficulty. Alternatively, terms including gradients of velocity and temperature may be incorporated into the equilibrium particle distribution, which is analogous to the formulation of Liu's modified countinous BGK equation[8]. Both methods can be applied to the lattice BGK model, though both of which have difficulties. First the use of collision matrix with different eigenvalues may lose the physically correct form for the viscous work in the macroscopic equation for the energy conservation[7,9]. On the other hand, gradient terms in the equilibrium particle distribution usually lower the efficiency of lattice BGK computation, because the absolute locality is destroyed in that information on the next nearest neighbours are required. However, the method introduced in this paper is both simple and effective. In the small temperature difference limit, the Prandtl number of the modeled fluid can vary in a relatively large range, comparing with the work of Bouchut and Perthame[6] on the continuous BGK model whose Prandtl number may vary only from 0 to 1. Application of this lattice BGK model to the more complicated Rayleigh-Benard flow is under current research work.

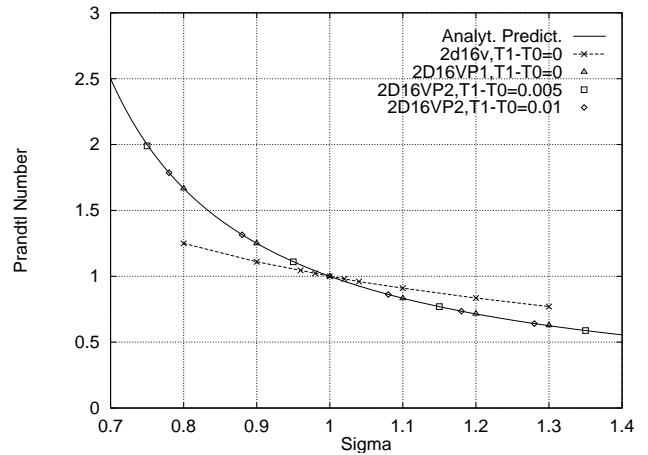

Fig. 1. Comparison of the numerical measured Prandtl numbers with the analytical prediction with formula (7). "2D16VP1" and "2D16VP2" are code names for different models. "2D16V" stands for two dimensions and sixteen velocity model, while "P1" represent a plain $\sigma$-incoporated model and "P2" a model with a further modification to minimize the deviation.

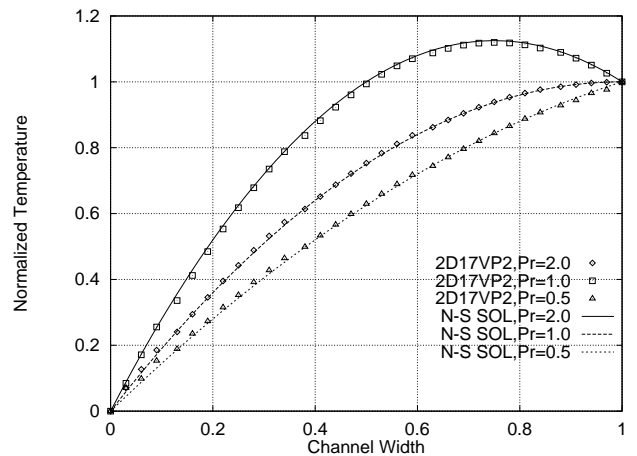

Fig. 2. Temperature pofiles obtained from lattice BGK simulation of the Couette flow. The lattice BGK model used here has 17 velocities, including the stationary particle distribution.